\documentclass{PoS}
\pdfoutput=1 
\usepackage{amsmath}
\usepackage{amssymb}
\usepackage{graphicx}
\usepackage{paralist}

\def\beq{\begin{equation}}
\def\eeq{\end{equation}}
\def\bea{\begin{eqnarray}}
\def\eea{\end{eqnarray}}

\newcommand{\Zp}{\ensuremath{Z^\prime} }

\newcommand{\SU}[1]{\ensuremath{\mathrm{SU}(#1)}}
\newcommand{\U}[1]{\ensuremath{\mathrm{U}(#1)}}

\title{NLO+NLL limits on $W'$ and $Z'$ gauge boson masses}

\ShortTitle{NLO+NLL limits on $W'$ and $Z'$ gauge boson masses}

\author{Tom\'{a}\v{s} Je\v{z}o$^a$, Michael Klasen$^b$, David~R.~Lamprea$^b$, \speaker{Florian Lyonnet}$^c$, Ingo Schienbein $^d$\\\\
\llap{$^a$} Universit\`a di Milano-Bicocca and INFN, Sezione di Milano-Bicocca,\\
			  Piazza della Scienza 3, 20126 Milano, Italy\\\\
\llap{$^b$}Institut f\"ur Theoretische Physik, Westf\"alische\\
				Wilhelms-Universit\"at M\"unster, Wilhelm-Klemm-Stra\ss{}e 9, D-48149\\
			  M\"unster, Germany\\\\
\llap{$^c$}Southern Methodist University, Dallas, TX 75275, USA\\\\
\llap{$^d$}Laboratoire de Physique Subatomique et de Cosmologie,\\
				Universit\'e Joseph Fourier/CNRS-IN2P3/ INPG,\\
				53 Avenue des Martyrs, F-38026 Grenoble, France\\\\
        E-mail: \email{tomas.jezo@mib.infn.it}, \email{michael.klasen@uni-muenster.de}, \email{david.lamprea@uni-muenster.de}, \email{flyonnet@mail.smu.edu}, \email{schien@lpsc.in2p3.fr}}

\abstract{QCD resummation predictions for the production of new charged ($W'$) and neutral
($Z'$) heavy gauge bosons decaying leptonically are presented. These results are obtained with our
resummation code at next-to-leading order and next-to-leading logarithmic (NLO+NLL) accuracy. Our predictions are compared to PYTHIA at leading order (LO) supplemented with parton showers (PS) and FEWZ at NLO and next-to-next-to-leading order (NNLO) for the $p_T$-differential and total cross sections in the Sequential Standard Model (SSM) and general SU(2)$\times$SU(2)$\times$U(1) models. We show that the importance of resummation for total cross sections increases with the gauge boson mass. Finally, the latest ATLAS and CMS results are reinterpreted to derive new limits at NLO+NLL on $W'$ and $Z'$ boson masses in general extensions of the Standard Model.}

\FullConference{The XXIII International Workshop on Deep Inelastic Scattering and Related Subjects\\
		April 27 - May 1, 2015\\
                Southern Methodist University\\
		Dallas, Texas 75275}

\begin{document}

\section{Introduction}
\label{sec:1}

New charged and neutral resonances are predicted in many well-motivated extensions of the Standard Model (SM) such as Grand Unified Theories (GUTs) or models with extra spatial dimensions. For various theoretical reasons (e.g.\ the hierarchy problem), new physics is expected to appear at the TeV scale and is searched for at the Large Hadron Collider (LHC), which is now operating at centre-of-mass energies of $\sqrt{S}= 13$ TeV (LHC13).

Experimental searches for $W'$ and $Z'$ bosons have so far mostly been performed in the Sequential Standard Model (SSM). While we adopt this model as a baseline to compare predictions with different theoretical accuracy, we also enlarge our analysis to a general $\text{G}(221) \equiv \text{SU}(2)_{1} \times \text{SU}(2)_{2}\times \text{U}(1)_{X}$ gauge group, which represents a well-motivated intermediate step towards the unification of the SM gauge groups.

Furthermore, the mass limits are mostly obtained using LO+PS Monte Carlo simulations with PYTHIA \cite{Sjostrand:2006za} rescaled to NNLO with FEWZ \cite{Gavin:2010az}, where neither of the programs include the a priori important interference effects of new and SM gauge boson exchanges.

In this paper, we present new QCD resummation predictions, which include these interference effects, at next-to-leading order and next-to-leading logarithmic (NLO+NLL) accuracy for the production of charged and neutral heavy gauge bosons ($W'$ and $Z'$) decaying into charged leptons and neutrinos. This calculation has been added to the list of already available processes in the publicly available NLO+NLL code RESUMMINO \cite{Fuks:2013vua}\footnote{Our code is available at \texttt{http://www.resummino.org}.}.


The results of our resummation code for $p_T$ distributions and total cross sections are compared using different benchmark models to  PYTHIA and FEWZ; for the sake of brevity we do not show our results on total cross sections. In addition, as an illustration we reinterpret the recent ATLAS $W'$ \cite{ATLAS:2014fk} and CMS $Z'$ \cite{CMS:2013qca} analyses using our NLO+NLL predictions, including interferences, and different new physics models.

\section{Set-Up and Numerical results}

We start by comparing numerical predictions obtained within the three different theoretical frameworks that we have developed/used: \begin{inparaenum}[(i)]\item the LO+PS Monte Carlo event generator PYTHIA in which we have implemented the full two to two process $pp\rightarrow W'/W\rightarrow \ell \nu$, \item the theoretical predictions in fixed order perturbation theory at NLO and NNLO QCD calculated with the FEWZ program, which unfortunately lacks the interference terms, \item our NLO+NLL QCD resummation implementation in RESUMMINO. \end{inparaenum} Details on our different implementations can be found in \cite{Jezo:2014wra}.

Apart from the SSM with identical fermion couplings of SM and new gauge bosons \cite{Altarelli:1989ff}, we also study the so-called G(221) models \cite{Hsieh:2010zr}, which are based on the intermediate semi-simple group $\SU{2}_1 \times \SU{2}_2 \times \U{1}_X$. In this framework, constraints on the parameter space from low-energy precision observables have been derived \cite{Hsieh:2010zr}, and several aspects of their phenomenology have already been studied \cite{Jezo:2012rm, Cao:2012ng,Abe:2012fb,Jinaru:2013eya,Jezo:2014kla}. Several well-known models emerge naturally from different ways of breaking the $\text{G}(221)$ symmetry down to the SM gauge group, in particular Left-Right (LR), Un-Unified (UU), Non-Universal (NU), Lepto-Phobic (LP), Hadro-Phobic (HP), and Fermio-Phobic (FP) models. 

In the following, only the results obtained in the UU realization will be presented even though our analysis also includes the NU model. Therefore, we list a subset of the five benchmark points ($B_{i}$, Model, $t$, $\Gamma_{W'}$ [GeV], $\Gamma_{W'\to\ell\nu}$ [GeV]) studied in \cite{Jezo:2014wra} for which results are shown here:\footnote{The PDFs are taken from the MSTW 2008 global fits at LO, NLO and NNLO, respectively. The renormalisation and factorisation scales $\mu_R$ and $\mu_F$ are identified with the new gauge boson mass $M_{V'}$, varied by a common factor of two up and down to estimate the scale uncertainty.} \begin{inparaenum}[(i)]\item $B_{1}$, SSM, --, 142.85, 11.69; \item $B_{2}$, UU, 0.7, 237.15, 5.73; \item $B_{3}$, UU, 1.2, 125.35, 16.83.\end{inparaenum}\ All the benchmark points are for a new gauge boson with a $4$ TeV mass.

	\subsection{Transverse momentum distributions}
%
\begin{figure}[t]
 \begin{center}
 \includegraphics[width=0.5\textwidth]{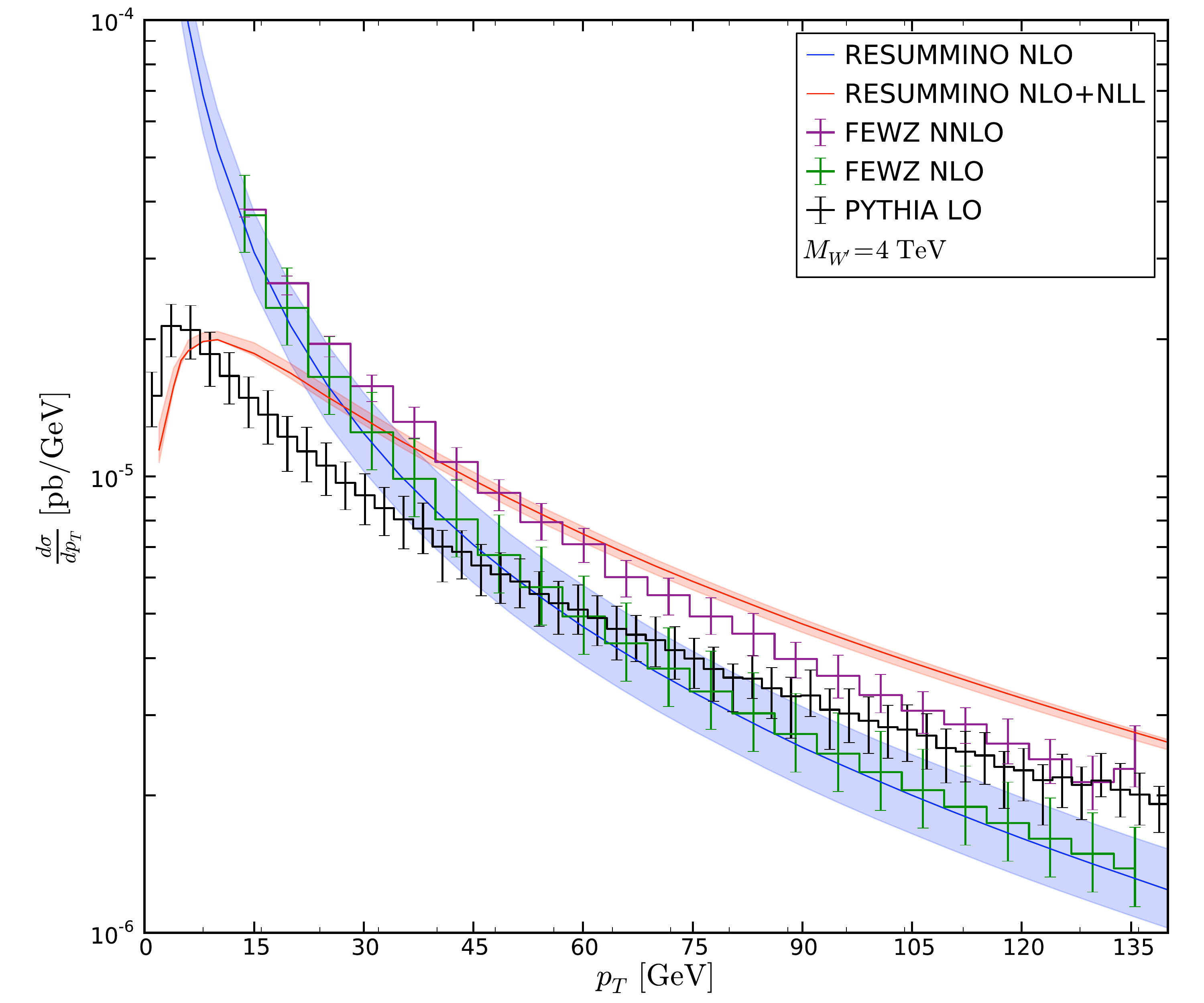}
 \end{center}
 \caption{Transverse momentum distributions of $W'$ bosons with a mass of 4 TeV at the
 LHC14 in the SSM. We compare our NLO and NLO+NLL predictions with RESUMMINO
 to those obtained with PYTHIA LO+PS, FEWZ NLO and NNLO.}
 \label{fig:2}
\end{figure}

Fig.\ \ref{fig:2} shows the $p_T$ spectrum for positively charged $W'$ bosons of mass 4 TeV
produced at LHC14 in the SSM and assumed to decay into a positron and an electron neutrino\footnote{A cut on the invariant mass of the lepton pair $Q>3M_{W'}/4$ is applied.}. The NLO predictions obtained with FEWZ and
RESUMMINO then agree very nicely, both for their central values and for their scale
uncertainties, and both diverge as $p_T\to0$. In contrast, the LO $\delta$-distribution
(not shown) is modified by the PYTHIA PS to a finite distribution, which exhibits a maximum around
$p_T\sim7$ GeV. A similar turnover, with a maximum at slightly larger values of $p_T\sim10$
GeV, is exhibited by the resummation calculation at NLO+NLL. The difference in shapes
can be attributed to different logarithmic accuracies (LL in PYTHIA, NLL in RESUMMINO),
while the one in normalisation comes mostly from the different perturbative orders
(LO in PYTHIA, NLO in RESUMMINO). At higher $p_T$ values, the NLO+NLL resummation
calculation agrees better with the fixed-order one by FEWZ at NNLO than at NLO, indicating
that important contributions beyond NLO are captured in the resummation approach. 

\subsection{Total cross sections and interferences}
\begin{figure}[h]
 \begin{center}
	\includegraphics[width=0.4\textwidth]{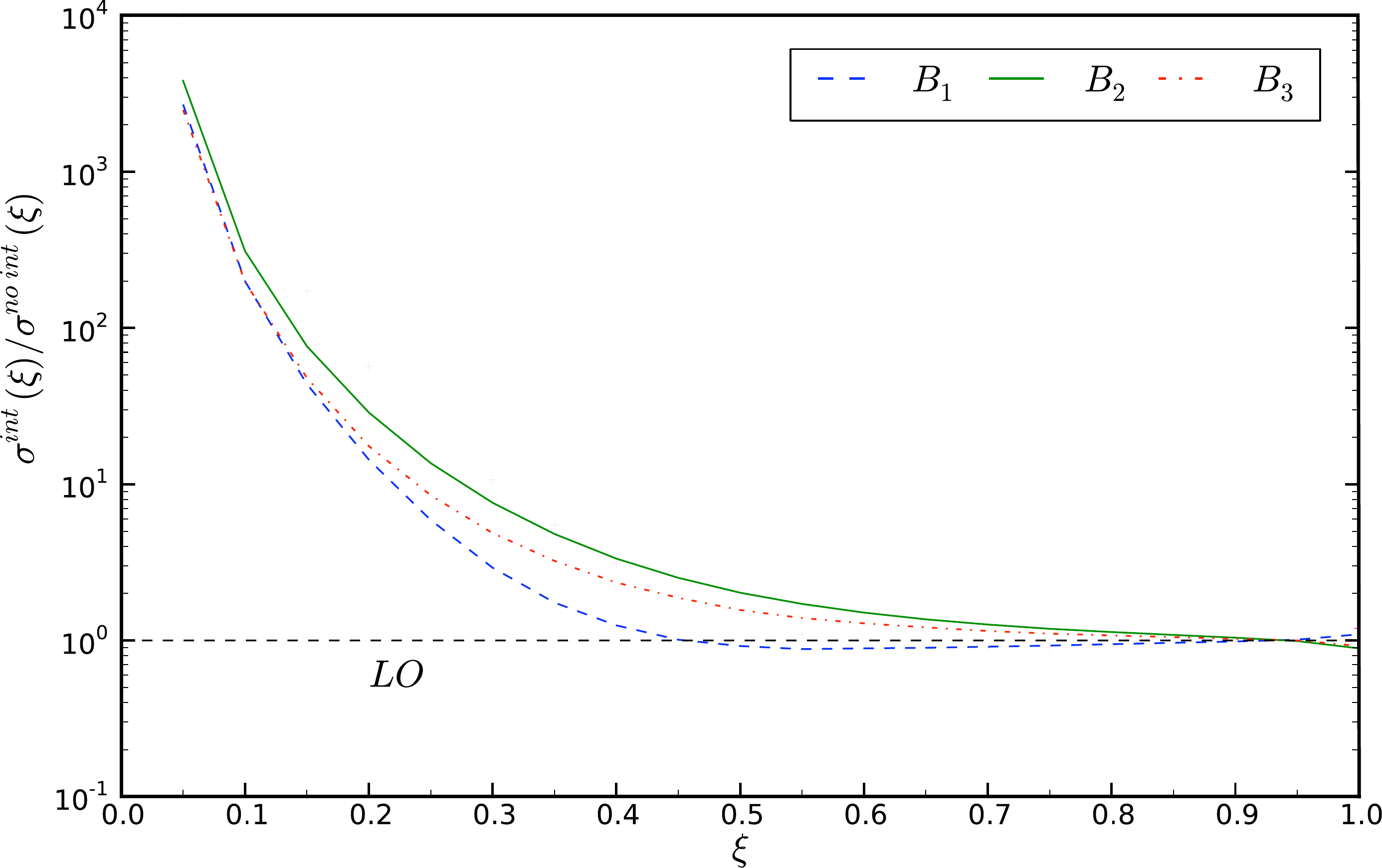}
	\includegraphics[width=0.34\textwidth]{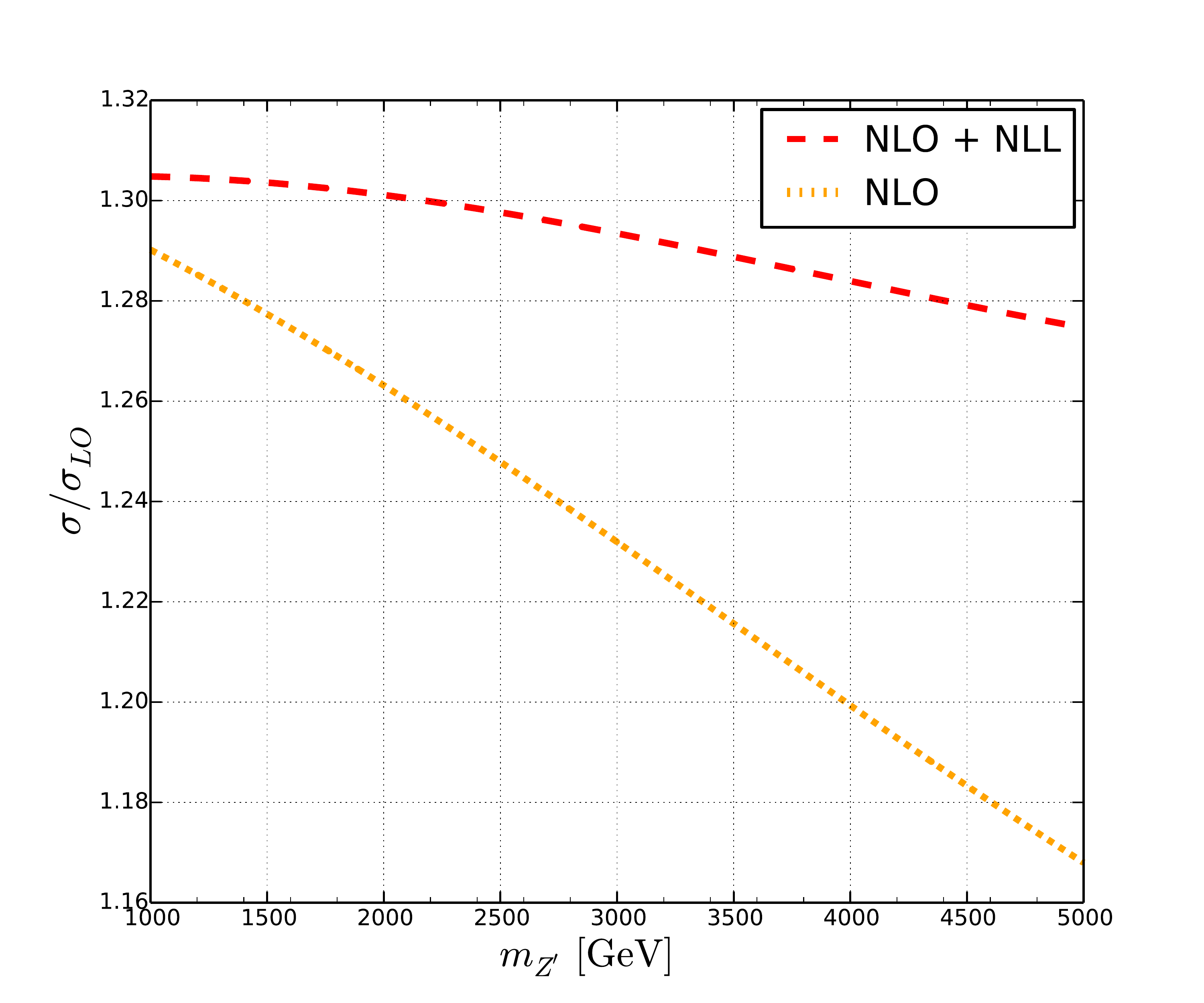}
 \end{center}
 \caption{Ratios of the total cross section at LO with and without interference
 terms as a function of the minimal invariant mass cut $Q>\xi M_{W'}$ for our three benchmark points $B_1,\ B_2,\ B_3$ (left). Ratios of $\Zp$ production cross sections at LHC14 at NLO and NLO+NLL over the LO cross section in the SSM and as a function of the heavy gauge boson mass (right).}
 \label{fig:3}
\end{figure}

The total cross sections for the five benchmark points have been compared in~\cite{Jezo:2014wra} and shown to agree within 1-2 percent for their central values and also, although somewhat less precisely, for their scale errors. The impact of the interference effect on these cross section is also important as one can see in Fig.\ \ref{fig:3} (left). Indeed, interference effects become quickly dominant as the invariant mass cut falls below 50\%. This fact will become important when we reanalyse the latest ATLAS and CMS results.

Finally, in Fig.\ \ref{fig:3} (right) the dependence of the
resummation contributions on the new gauge boson mass, using now the example
of a neutral $Z'$ gauge boson is shown.  
A look at the NLO+NLL prediction shows that as one approaches the threshold region the resummation
of logarithms becomes increasingly important, making our resummation calculations even more relevant as the LHC explores
higher and higher mass regions.

\section{Gauge boson mass limits in general SM extensions}
\label{sec:4}

In this section, we reanalyse the latest experimental searches by the ATLAS and CMS
collaborations for $W'$ and $Z'$ bosons in their leptonic decay channels, performed at LHC8 in
the SSM. We use our resummation predictions at NLO+NLL and do this not only in the SSM,
but also in the UU and NU (not shown) models that have previously not been considered. 

\subsection{ATLAS limits on $W'$ boson masses}

\begin{figure}[b!]
 \centering
 \includegraphics[width=0.45\textwidth]{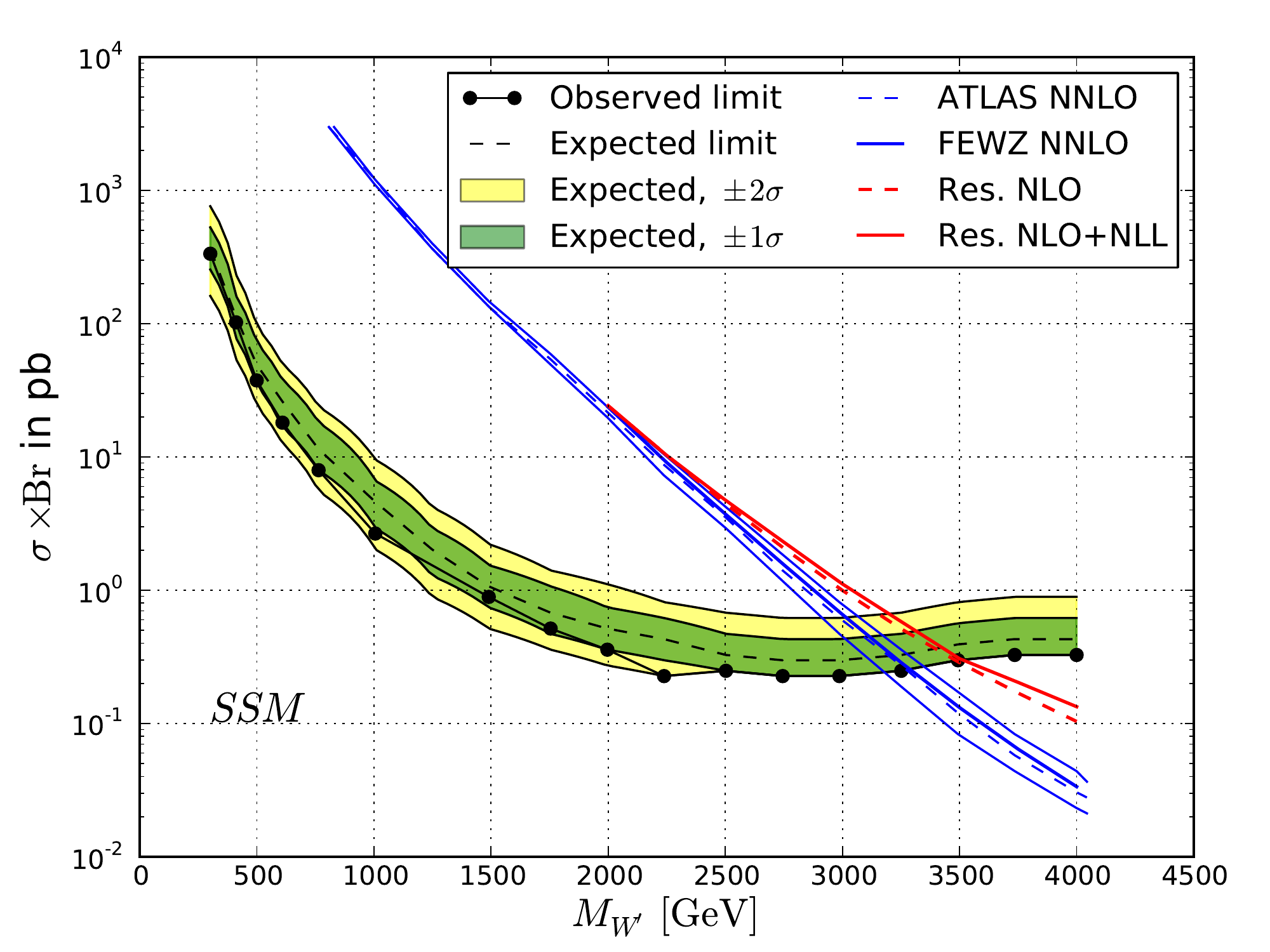}
 \includegraphics[width=0.45\textwidth]{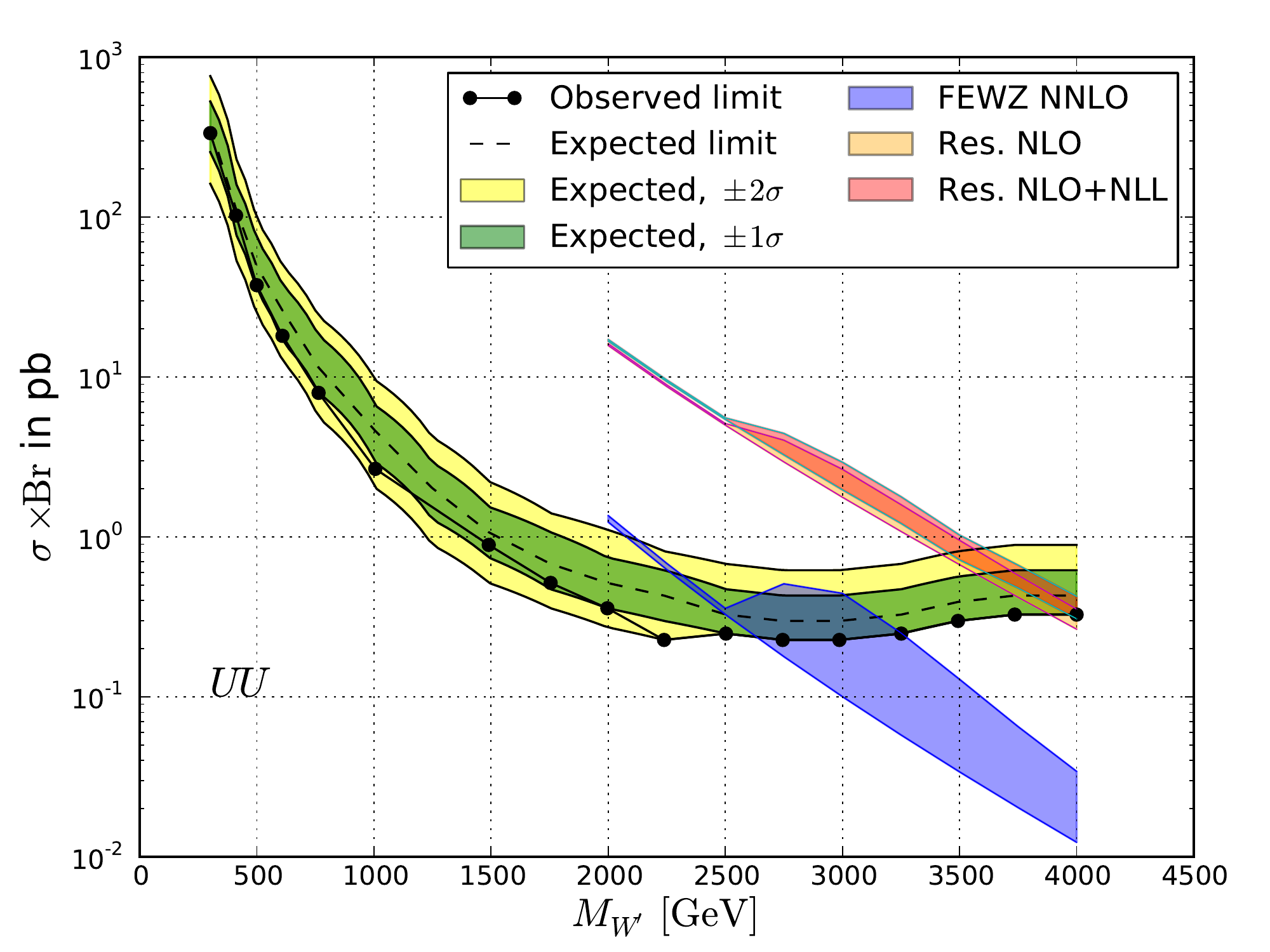}
 \caption{Left: Cross sections times branching ratios for SSM $W'$ bosons decaying into electrons or muons
 and neutrinos at LHC8. The expected (dashed black) and observed (full black) limits in the ATLAS
 analysis \cite{ATLAS:2014fk}, using a cut of $Q>0.4M_{W'}$ at the generator level, and their corresponding
 uncertainties at the 1$\sigma$ (green) and 2$\sigma$ (yellow) level are compared to predictions
 without interference at NNLO in ZWPROD and in
 FEWZ (central only, full blue) and with interference at NLO (central only, dashed red) and at NLO+NLL (central only,
 full red) using RESUMMINO. Right: Same as left figure for UU model $W'$ bosons.}
  \label{fig:5}
\end{figure}
%
The ATLAS analysis is performed by simulating the $W'$ signal with PYTHIA LO+PS, adding negative and positive charges, and rescaling it to NNLO total cross section accuracy with the program ZWPROD. As a consequence, the interference effects between SM $W$ bosons and SSM $W'$ bosons are not included. Looking at the RESUMMINO predictions that include interference effects, these are seen to be very important, since the invariant mass cut is relatively low (cf.\ Fig.\ \ref{fig:3} (left)), and they lead to an increase of $\sigma\times$Br of about a factor of two at the highest mass considered here. There, the resummation effects are also best visible, and they increase the NLO prediction (dashed red) by about 20\% at NLO+NLL (full red). This leads to a lower bound on $W'$ bosons masses of 3.5 TeV, see Fig.~\ref{fig:5} (left).
%
\begin{figure}[t]
 \centering
  \includegraphics[width=0.45\textwidth]{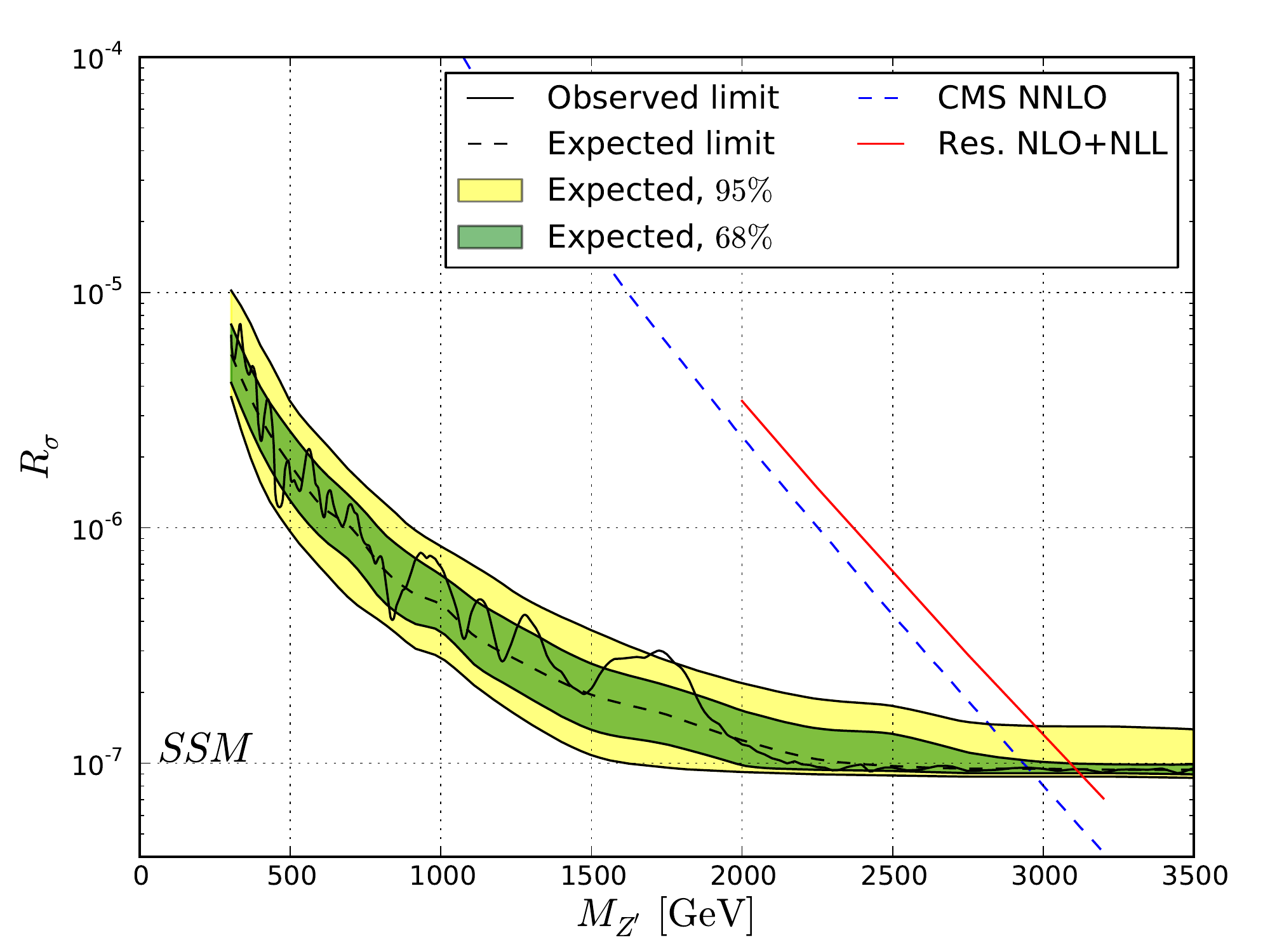}
  \includegraphics[width=0.45\textwidth]{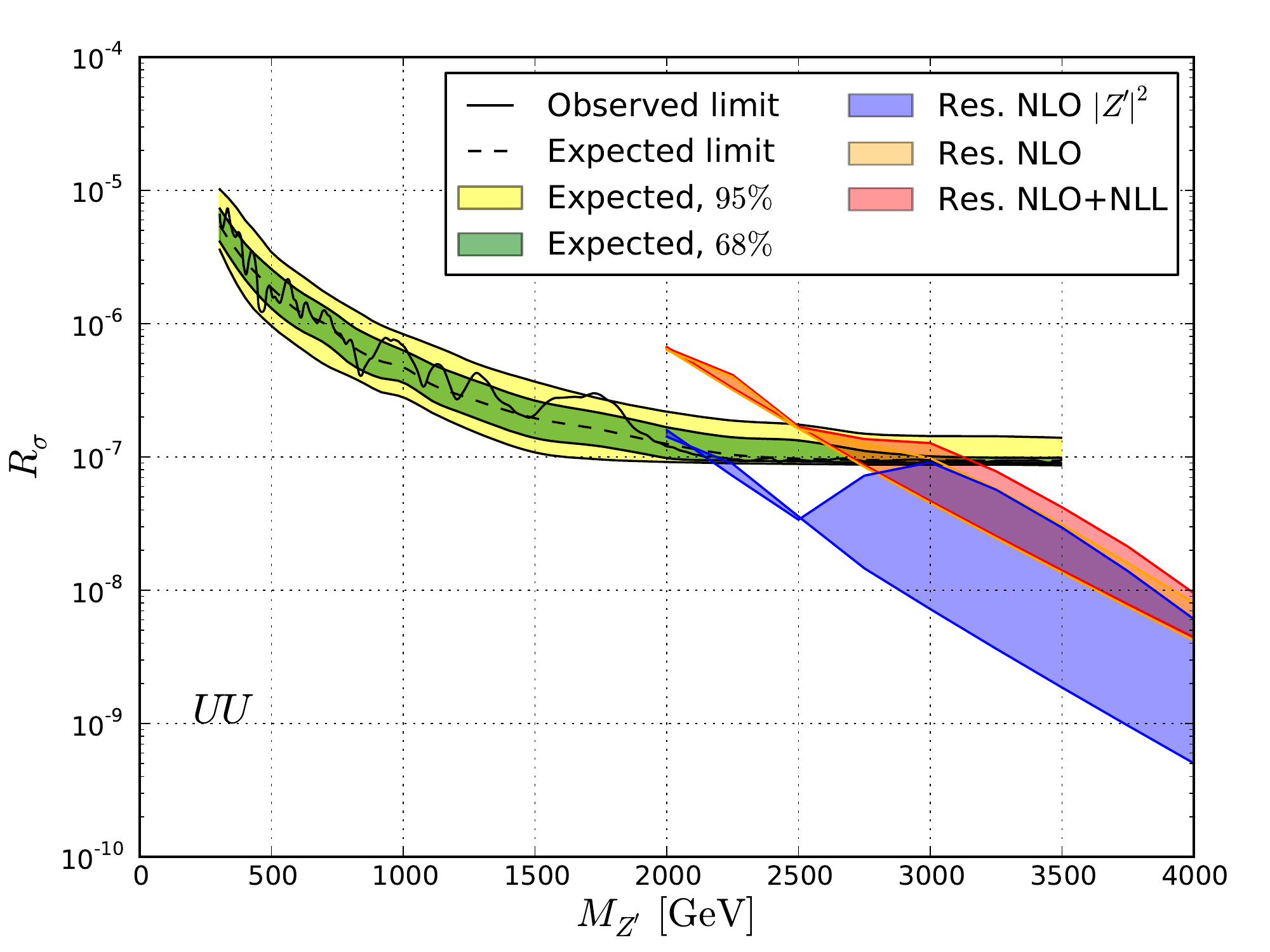}
 \caption{Left: Ratios of new physics over SM cross sections for SSM $Z'$ bosons decaying into electron or muon
 pairs at LHC8. The expected (dashed black) and observed (full black) limits in the final CMS
 analysis \cite{CMS:2013qca}, using a cut of $0.6M_{Z'}<Q<1.4M_{Z'}$, and their corresponding
 uncertainties at the 68\% (green) and 95\% (yellow) C.L.\ are compared to predictions
 without photon, $Z$ and $Z'$ interference at NNLO in ZWPROD (dashed blue)
 and with full interference at NLO+NLL (full red) using RESUMMINO. Right: Same as left figure for UU model $Z'$ bosons.}
  \label{fig:6}
\end{figure}

The results for the UU model are presented in Fig.\ \ref{fig:5} (right). For each $W'$ boson mass, the variation of the $t$ parameter
in the allowed range leads to a spread of theoretical predictions\footnote{Below masses of 2.5 TeV, where the UU model is already excluded by low-energy and precision constraints \cite{Hsieh:2010zr}, the areas have been shrunk to a single line, calculated for a hypothetical $t$-value of 0.18 pertinent at the same time to the minimal allowed mass and the perturbativity limit.}. The inclusion of interference effects in RESUMMINO leads to an increase of the predicted cross sections by almost an order of magnitude at $M_{W'}=4$ TeV. At NLO+NLL and including interference effects, our reanalysis excludes $W'$ bosons in
the UU model with masses below 3.9--4 TeV, which considerably improves the limits from low-energy
and precision constraints\footnote{As in this model $M_{Z'}\simeq M_{W'}$ up to corrections of ${\cal O}
(v^2/u^2)$, this implies an identical mass limit for $Z'$ bosons in the UU model.}.

Our analysis in the NU model is not shown here but can be found in \cite{Jezo:2014wra}.

\subsection{CMS limits on $Z'$ boson masses}

The CMS collaboration have searched for narrow resonances in the dilepton
(electron or muon) mass spectrum and set mass limits of 2.96 TeV and 2.6 TeV
on SSM $Z'$ bosons and a specific class of superstring-inspired $Z'$ bosons,
respectively \cite{CMS:2013qca}.

The mass limits are obtained by comparing expected and observed experimental
limits on $R_\sigma$ with expectations from PYTHIA LO+PS, rescaled to NNLO
with ZWPROD. For the SSM, we show the result in Fig.\ \ref{fig:6} (left), where one
can read off the limit cited above. The inclusion of the interferences leads to a considerable
increase of the prediction, computed by us with RESUMMINO at NLO+NLL, so that the SSM
exclusion limit moves to 3.2 TeV.

For $Z'$ bosons in the UU model, we simulate in Fig.\ \ref{fig:6} (right) the ratio $R_\sigma$. Interference effects increase again the prediction (light red) by about an order of magnitude, while the additional radiative corrections at NLO+NLL (dark red) do not alter the result significantly. In the UU model, we then obtain $Z'$ boson mass limits ranging from 2.75 TeV up to 3.2 TeV. These are in all cases stronger than the previously obtained indirect limit of 2.5 TeV.

Again, similar results have been obtained for the NU model, see \cite{Jezo:2014wra}.

\section{Conclusion}
\label{sec:5}

In this contribution, we have presented resummation calculations at NLO+NLL accuracy for the production of leptonically decaying $W'$ and $Z'$ bosons in hadronic collisions at small transverse momenta and/or close to production threshold. Our calculations include the full interference structure of new and SM gauge bosons and therefore currently provide the best available theoretical precision for realistic cross section estimates. To facilitate a comparison with LO+PS calculations, we furthermore implemented interference effects in PYTHIA by adding a new $2\to2$ process, i.e.\ without relying only on resonant production or the narrow width approximation. 

We presented a detailed numerical analysis of the predictions obtained within three different frameworks and showed that the agreement was at the 1-2\% level on the total cross sections where expected. The various predictions for the $p_T$ spectrum were also compared in great detail. Finally, it was shown that the intereferences play a crucial role even when kinematical cuts are placed and that the importance of resummation grows with increasing mass of the heavy gauge boson.

Through a reanalysis of the currently strongest ATLAS exclusion limits of $W'$ boson masses in the SSM, we showed that $W'$ boson masses could be excluded below 3.9--4 TeV in the UU model. Similarly, a reanalysis of the currently strongest CMS exclusion limits of $Z'$ boson masses in the SSM led to an exclusion limit of 2.75--3.2 TeV, which was again stronger in the UU model than the low-energy and precision constraints of 2.5 TeV. 

\clearpage

\bibliographystyle{JHEP}
\bibliography{resummation_proceeding.bbl}

%
%

\end{document}